\documentclass[a4paper,11pt]{article}
\usepackage{pos}
\usepackage{natbib}
\usepackage{soul}

\title{Evidence for proton acceleration and escape from the Puppis A SNR using Fermi-LAT observations}
 \ShortTitle{Gamma-ray emission from Puppis A}

\author*[a,b]{Giuffrida, R.}
\author[c]{Lemoine-Goumard, M.}
\author[a,b]{Miceli, M.}
\author[d]{Gabici, S.}
\author[e,f]{Fukui, Y.}
\author[g]{Sano, H.}

\affiliation[a]{Dipartimento di Fisica e Chimica E. Segr\`e, Universit\`a degli Studi di Palermo, Piazza del Parlamento 1, 90134, Palermo, Italy}

\affiliation[b]{ INAF-Osservatorio Astronomico di Palermo, Piazza del Parlamento 1, 90134, Palermo, Italy}

\affiliation[c]{Université Bordeaux, CNRS/IN2P3, LP2I Bordeaux, UMR 5797, F-33170 Gradignan, France}
\affiliation[d]{Université de Paris, CNRS, Astroparticule et Cosmologie, F-75013 Paris, France}
\affiliation[e]{Department of Physics, Nagoya University, Chikusa-ku, Nagoya 464-8602, Japan}
\affiliation[f]{Institute for Advanced Research, Nagoya University, Furo-cho, Chikusa-ku, Nagoya 464-8601, Japan
}

\affiliation[g]{Faculty of Engineering, Gifu University, 1-1 Yanagido, Gifu 501-1193, Japan}

\emailAdd{roberta.giuffrida@inaf.it}

\abstract{Supernova remnants (SNRs) are the best candidates for galactic cosmic ray acceleration to relativistic energies via diffusive shock acceleration. The gamma-ray emission of SNRs can provide direct evidence of leptonic (inverse Compton and bremsstrahlung) and hadronic (proton-proton interaction and subsequently pion decay) processes. Puppis A is a $\sim$4 kyr old SNR interacting with interstellar clouds which has been observed in a broad energy band, from radio to gamma-ray. We performed a morphological and spectral analysis of 14 years of observations with \textit{Fermi}-LAT telescope in order to study its gamma-ray emission. We found a clear asymmetry in high-energy brightness between the eastern and western sides of the remnant, reminiscent to that observed in the X-ray emission. The eastern side, interacting with a molecular cloud, shows a spectrum which can be reproduced by a pion decay model. Moreover, we analyzed two gamma-ray sources located close to the remnant. The hardness of their spectra suggests that the gamma-ray emission can be due to particles escaping from the shock of Puppis A.}

\ConferenceLogo{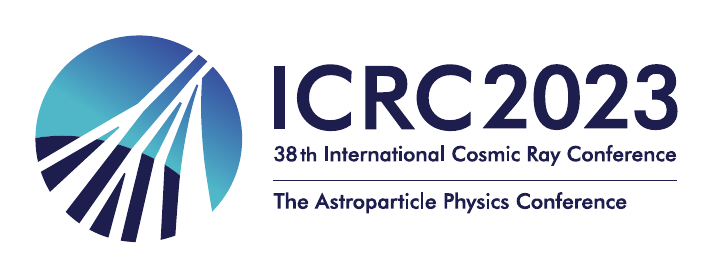}

\FullConference{%
38th International Cosmic Ray Conference (ICRC2023)\\
  26 July - 3 August, 2023\\
  Nagoya, Japan}

\begin{document}
\maketitle
\section{Introduction}
Supernova Remnants (SNRs) are the primary suspect among Galactic sources to accelerate particles via diffusive shock acceleration up to the necessary PeV energies \cite{bel78}. Since the supernova explosion releases $\sim 10^{51}$ erg, considering 2 supernova explosions per century in the Milky Way \cite{bvc21} and the requested cosmic ray power (of the order of $10^{41}$ erg s$^{-1}$, SNRs have to transfer about 10\% of their kinetic energy in order to accelerate particles \cite{hil05}). 
Cosmic ray acceleration from SNRs can be investigated with the gamma-ray emission from the remnant. Both leptonic and hadronic acceleration produce gamma-ray emission from the remnant via different mechanism: inverse Compton (IC) and/or bremsstrahlung for the leptonic scenario and pion decay for the hadronic scenario. In particular, the proton-proton interaction can be enhanced in SNRs evolving in a dense interstellar medium, such as molecular clouds, where the particles accelerated at the shock front interact with the dense gas.\\
Puppis A is an interesting source to study particle acceleration because of its interaction with interstellar clouds \cite{da88,asf22}. It is a 4 kyr old \cite{wk85,bpw12}, shell-like supernova remnant at a distance of about 1.3 kpc \cite{rcw17} with a 56' projected diameter. Its high energy emission has been observed in different energy bands such as radio, X-rays and gamma-rays. 
The main feature of the X-ray emission, the so called Bright Eastern Knot (BEK), shows presence of shock-cloud interaction in the eastern side of Puppis A \cite{hfp05}. The asymmetry in the X-ray emission \cite{dlr13} shows that the remnant is evolving in an inhomogeneous ambient medium. Similar results have been obtained through the analysis of \textit{eROSITA} data, which also showed that the southwester part of the remnant is subject to strong foreground absorption \cite{mbp22}.\\
The gamma-ray emission from Puppis A has been already analyzed \cite{hgl2012,xgl17} with \textit{Fermi}-LAT telescope, revealing an asymmetry in the morphology of the source between its eastern and western side. The two parts have been analyzed separately in 200 MeV - 100 GeV energy band \cite{hgl2012} showing the spectrum of the western side is steeper by an index of 0.3 with respect to the spectrum of the eastern side. Nevertheless, the low statistic was not sufficient to constrain the process responsible of the gamma ray emission. In conclusion, the total energy of accelerated particles results in $5 \times 10^{49}$ erg \cite{hgl2012}.
The very high energy emission from Puppis A has been investigated with H.E.S.S. \cite{HessColl15}, resulting a non detection which can be signature of particle acceleration ceased some time ago. 
We analyzed the gamma-ray emission from Puppis with \textit{Fermi}-LAT telescope and we can confirm the asymmetry in the morphology of the remnant which can be well reproduced by its soft X-ray emission. \\
Moreover, we analyzed two different gamma-ray excesses close to the remnant, 4FGL J0822.8-4207 and PS J0824.0-4329, whose spectral shapes suggest the origin of gamma ray emission due to cosmic rays acceleration and escape from Puppis A.

\section{Data Analysis and Results}
We analyzed \texttt{P8R3} data  with 14 years of observations with \textit{Fermi}-LAT telescope in the region of interest (ROI) within 15° around the source 4FGL J0822.1-4253e (Puppis A) using the Fourth Source Catalog (4FGL-DR3 \cite{4fgl2020,DR32022}). Data have been analyzed with the software \texttt{fermitools\footnote{https://fermi.gsfc.nasa.gov/ssc/data/analysis/software/}} version 2.2.0 and the \texttt{PYTHON} package \texttt{fermipy\footnote{https://fermipy.readthedocs.io/en/latest/}} version 1.2 and filtered with \texttt{DATA\_QUAL>0, LAT\_CONFIG==1}. 
All the source in the catalog, the Galactic diffuse emission and the instrument response functions (IRFs) are taken into account to model the background.

\subsection{Morphological analysis}
\label{sec:morph}
\begin{figure}
    \centering
    {\includegraphics[scale=0.4]{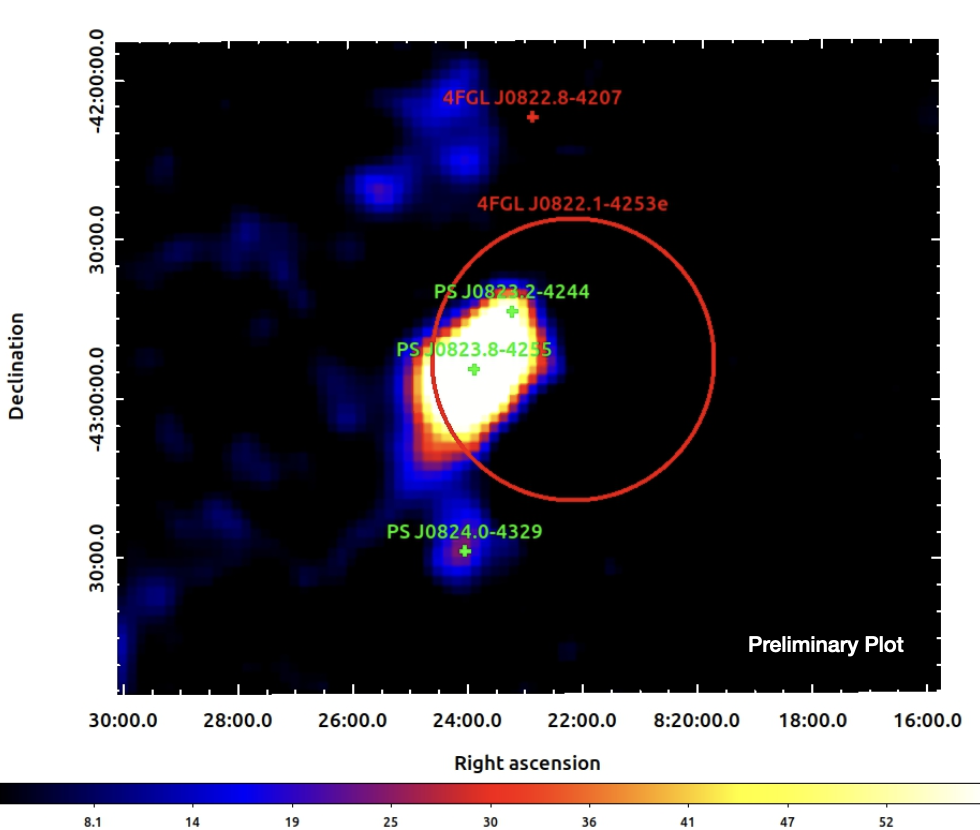}}
    {\includegraphics[scale=0.4]{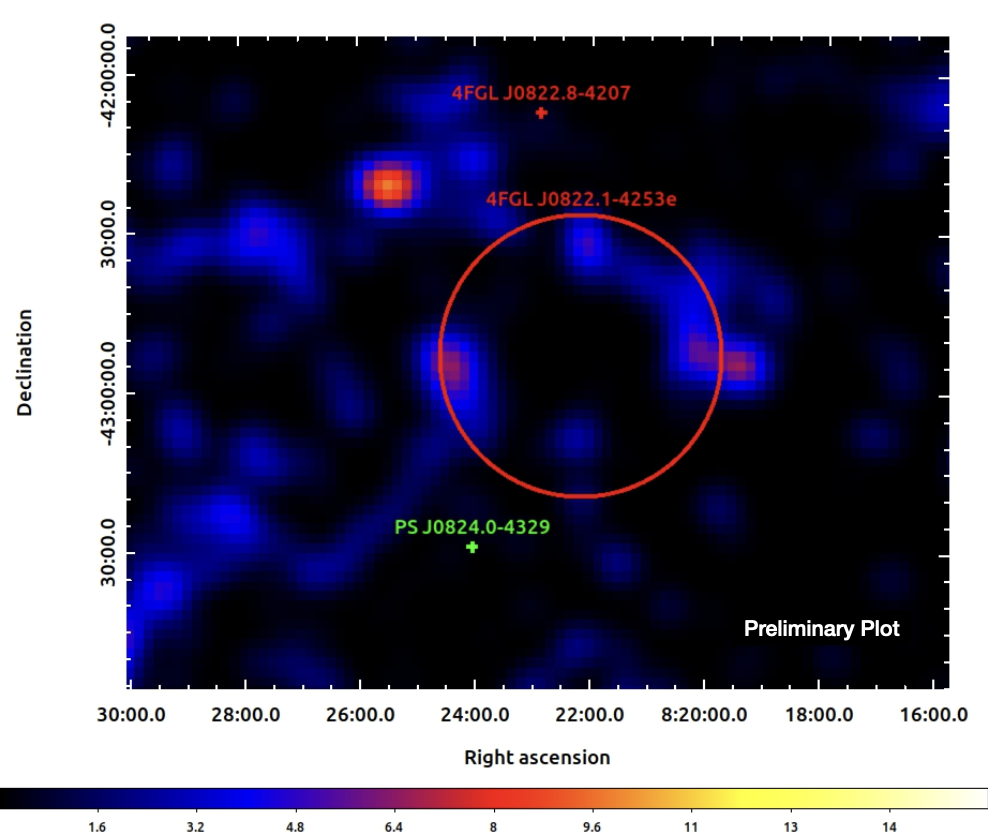}}
    \caption{Preliminary plots. Left panel: residual TS map obtained by fitting the ROI (see text) with the catalog 4FGL-DR3. The red circle shows the uniform disk in 4FGL-DR3 which models Puppis A, the red cross shows the source 4FGL-J0822.8-4207 and the green crosses indicate the new sources needed to fit the gamma ray excesses (Tab. \ref{tab:PS}). Right panel: residual TS map obtained using the \textit{XMM-Newton} observations in 0.7-1.0 keV energy band corrected for the $N_H$ to model Puppis A. The red circle, red cross and green cross are the same as in the left panel.}
    \label{fig:Resid}
\end{figure}

Data have been analyzed in 1-1000 GeV energy band in order to have a good angular resolution to reproduce the morphology of the gamma-ray emission from Puppis A. 
The left panel of Fig.\ref{fig:Resid} shows the residual TS map obtained using the catalog 4FGL-DR3 where Puppis A is fitted as a uniform disk. The map shows a bright excess in the eastern part of the remnant, indicating an asymmetry in the gamma-ray emission which is not well reproduced by the uniform disk. The fit improves by adding two point-like sources \cite{agk2022}: PS J0823.2-4244 and PS J0823.8-4255 (Tab. \ref{tab:PS}).
Moreover, we also found a gamma-ray excess in the south of the remnant, which can be well fitted with a point-like source (PS J0824.0-4329) with a TS value of 18.5. 
\begin{table}
	\caption{Coordinates of additional point-like sources (see the text).}
	\begin{center}
		\footnotesize
		\begin{tabular}{c c c}
		    \hline\hline
			Source & RA (degrees) & DEC (degrees) \\
			\hline
		    PS J0823.2-4244 & 125.8 & -42.7\\
            PS J0823.8-4255 & 126.0 & -42.9 \\
            PS J0824.0-4329 & 126.0 & -43.5 \\
            \hline
		\end{tabular}
		\label{tab:PS}
	\end{center}
\end{table}
It has been found \cite{hgl2012} that the gamma-ray emission of Puppis A can be very well reproduced with the X-ray template obtained by \textit{ROSAT} observations.
We here consider the more recent X-ray observations performed with \textit{XMM-Newton} and inspect different energy bands. 
Indeed, the X-ray emission also shows a difference in fluxes between East and West sides of the remnant. On the other hand, this asymmetry is not visible in the radio band.  
We compared different templates with the Akaike Information Criterion (AIC) \cite{Akaike1998} which allows us to compare even not nested templates basically with the AIC value, given by  $AIC = -2\ln(L) + 2k$, where $k$ is the number of degree of freedom and L the likelihood. 
We found that the soft X-ray band ($0.7-1.0$ keV) provides a significant improvement of the fit with respect to the hard X-ray band ($1.0-8.0$ keV). However, the effects of the absorption need to be included in the soft X-ray analysis. We then corrected the observed maps for the (non-uniform) interstellar absorption, and produced unabsorbed X-ray maps. For each point of the X-ray count-rate map, we obtained the corresponding unabsorbed count-rate by assuming the local value of $N_H$ derived by the spatially resolved spectral analysis \cite{lsd16}. This unabsorbed X-ray map in the 0.7-1.0 keV band provides the best fit to the gamma-ray data (Tab .\ref{tab:template}).
The right panel of the Fig. \ref{fig:Resid} shows the residual TS map obtained fitting the source with this map. Even with this template, we confirm the presence of the point-like source PS J0824.0-4329 and of the already studied source 4FGL J0822.8-4207 \cite{agk2022}.
\begin{table}
	\caption{Comparison between different templates used to reproduce the morphology of Puppis A}
	\begin{center}
		\footnotesize
		\begin{tabular}{c c c c}
		    \hline\hline
			Template & Log Likelihood & d.o.f. & $\Delta$AIC \\
			\hline
            Catalog + 2pt sources & 609.1 & 14 & 0 \\
            \textit{XMM-Newton} (1.0-8.0 keV) & 600.1 & 3 & 4\\
            \textit{XMM-Newton} (0.7-1.0 keV) & 618.9 & 3 & 41 \\            \textit{XMM-Newton} (0.7-1.0 keV) corrected for the $N_H$ & 624.0 & 3 & 52\\
            2 half \textit{XMM-Newton} (0.7-1.0 keV) corrected for the $N_H$ & 624.8 & 6 & 47\\
            \hline
		\end{tabular}
		\label{tab:template}
	\end{center}
\end{table}
We then analyzed the source 4FGL J0822.8-4207 \cite{agk2022} (TS=53.6) in order to check if it is a point-like or an extended source (we considered a Gaussian for the spatial model of the extended case). Using the fermipy tool to quantify the extension significance, it shows a TS$_{ext}$= 22.0, with TS$_{ext}$ = 2 $\ln L_{ext}/L_{ps}$ where $L_{ext}$ and $L_{ps}$ are the likelihoods for the fit with the extended and the point-like source, respectively. As a conclusion, with 14 years of observations 4FGL J0822.8-4207 satisfied the rules for extended sources (TS$_{ext}>16$) \cite{laa2012}, showing that it can be well fitted as a gaussian with sigma 0.15 deg centered at ra= 125 deg, dec= -42.3 deg.
The same method has been followed for the excess in the south whose TS$_{ext}=4$ does not reach the minimum requested value of 16 to be considered as extended. 

\subsection{Spectral analysis}
\label{sec:spec}
\begin{figure}
    \centering
    \includegraphics[width=0.5\textwidth]{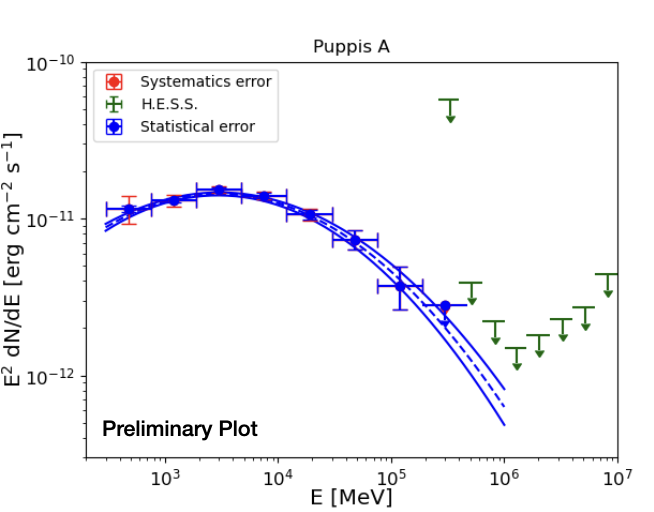}
    \caption{Blue: SED obtained for Puppis A in 300 MeV - 1 TeV energy band. The spectrum has been fitted by the model in Eq. \ref{Eq:LogP}. Red: upper limits obtained with H.E.S.S \cite{HessColl15}. Statistical and systematics error are shown in blue and in red respectively.}
    \label{fig:pup}
\end{figure}

The spectral analysis has been performed in 300 MeV - 1 TeV energy band using the best template found in Sec. \ref{sec:morph}. 
The spectrum of Puppis A (Fig. \ref{fig:pup}) has been fitted using the spectral model \texttt{Log Parabola} in \texttt{gtlike}:

\begin{equation}
    \frac{dN}{dE} = N_0 \left(\frac{E}{E_b}\right)^{-(\alpha + \beta ln(E/E_b))}.
    \label{Eq:LogP}
\end{equation}

The new \textit{Fermi}-LAT SED, combined with the H.E.S.S. upper limits \cite{HessColl15} is shown in Fig.\ref{fig:pup}.

The two sources 4FGL J0822.8-4207 and PS J0824.0-4329 have been fitted with a power law model. Since both sources are located in molecular clouds \cite{asf22}, their gamma-ray flux could be due to hadronic emission from particles accelerated at the shock of Puppis A. 
4FGL J0822.8-4207 shows an hard spectrum with a photon index of $2.07\pm0.08$ in 300 MeV - 1 TeV energy band (Tab. \ref{tab:fit}). 

Similarly, although the faint emission of the source in the south can not allow us to claim it as a significant source (TS value = 19, significance of less than 5 $\sigma$), it also takes place in a dense ambient medium and presents an hard spectrum with photon index of $1.7\pm0.2$. More statistics can provide interesting information on the source PS J0824.0-4329.

In order to study the asymmetry between the eastern and western side of Puppis A, we analyze the spectra of the two parts separately. In both cases the \texttt{LogParabola} (Eq. \ref{Eq:LogP}) provides the best fit and, in particular, both spectra seem to reproduce the hadronic scenario. 
The origin of the emission (acceleration from the thermal gas or reacceleration of existing cosmic rays) still needs to be investigated. 
The best fit parameters for the studied sources are summarized in Tab. \ref{tab:fit}.

\begin{table}
	\caption{Best fit values for the analyzed sources in 300 MeV - 1 TeV energy band. Only statistical errors are provided. The prefactor is taken at the reference energy of 1 GeV.}
	\begin{center}
		\footnotesize
		\begin{tabular}{c c c c c c}
		    \hline\hline
			Source & Prefactor (cm$^{-2}$ s$^{-1}$) & Photon Index & alpha & beta & TS value \\
			\hline
            4FGL J0822.8-4207 & $(2.8 \pm 0.3) \times 10^{-14}$& 2.07 $\pm$ 0.08 & - & - & 19\\
            PS J0824.0-4329 & $(6 \pm 5) \times 10^{-14}$& 1.7 $\pm$ 0.2 & - & - & 98\\
            Puppis A & $(8.0 \pm 0.2) \times 10^{-12}$ & - & 1.80 $\pm$ 0.03 & 0.09 $\pm$ 0.01 $\pm$ 0.02 & 5474\\
            Puppis A East & $(5.3 \pm 0.2)\times 10^{-12}$ & - & 1.80 $\pm$ 0.04 & 0.09 $\pm$ 0.01 & 2801\\
            Puppis A West & $(2.7 \pm 0.2)\times 10^{-12}$ & - & 1.8 $\pm$ 0.1 & 0.10 $\pm$ 0.05 & 717\\
            \hline
		\end{tabular}
		\label{tab:fit}
	\end{center}
\end{table}

\section{Summary and conclusion}
We analyzed 14 years of observations with \textit{Fermi}-LAT telescope in order to study the gamma-ray emission from Puppis A. The morphological analysis of the remnant, in 1-1000 GeV energy band, have confirmed the asymmetric emission between East and West side \cite{hgl2012} which echoes the same asymmetry in its X-ray emission \cite{dlr13}. In particular, the best template which reproduces the gamma-ray emission from Puppis A is provided by the \textit{XMM-Newton} observation in the soft X-ray band (0.7-1.0 keV) corrected for the variable $N_H$ \cite{lsd16} (Tab. \ref{tab:template}). The spectral analysis shows that the two sides have compatible spectral shapes. Moreover, we analyzed two gamma ray excesses (4FGL J0822.8-4207 and PS J0824.0-4329) outside Puppis A that appear close to molecular emission and show harder spectra than Puppis A above 10 GeV.

We plan to investigate in detail the physical origin of the gamma-ray emission of Puppis A.We also note that the curved spectral shape that we observe is consistent with the lack of emission at very high energies.

%
%
%
\acknowledgments
The \textit{Fermi}-LAT Collaboration acknowledges support for LAT
development, operation and data analysis from NASA and DOE (United
States), CEA/Irfu and IN2P3/CNRS (France), ASI and INFN (Italy), MEXT,
KEK, and JAXA (Japan), and the K.A.~Wallenberg Foundation, the Swedish
Research Council and the National Space Board (Sweden). Science analysis
support in the operations phase from INAF (Italy) and CNES (France) is
also gratefully acknowledged. This work performed in part under DOE
Contract DE-AC02-76SF00515.\\
MLG acknowledges support from Agence Nationale de la Recherche (grant
ANR- 17-CE31-0014).
RG and MM acknowledge support from the INAF mini-grant ``X-raying shock 
modification in supernova remnants''
\bibliographystyle{JHEP}
\bibliography{references}
\end{document}